\documentclass[sigconf]{acmart}

\AtBeginDocument{%
  \providecommand\BibTeX{{%
    \normalfont B\kern-0.5em{\scshape i\kern-0.25em b}\kern-0.8em\TeX}}}

\copyrightyear{2020} 
\acmYear{2020} 
\setcopyright{rightsretained} 
\acmConference[EICC 2020]{European Interdisciplinary Cybersecurity Conference}{November 18, 2020}{Rennes, France}
\acmBooktitle{European Interdisciplinary Cybersecurity Conference (EICC 2020), November 18, 2020, Rennes, France}
\acmDOI{10.1145/3424954.3424968}
\acmISBN{978-1-4503-7599-3/20/11}

\begin{document}

\title[Towards Reconstructing Multi-Step Cyber Attacks in Modern Cloud Environments with Tripwires]%
  {Towards Reconstructing Multi-Step Cyber Attacks\texorpdfstring{\\}{ }in Modern Cloud Environments with Tripwires}

\author{Mario Kahlhofer}
\email{mario.kahlhofer@dynatrace.com}
\orcid{0000-0002-6820-4953}
\affiliation{%
  \institution{Dynatrace Research}
  \city{Linz}
  \country{Austria}
}

\author{Michael H{\"o}lzl}
\email{michael.hoelzl@dynatrace.com}
\orcid{0000-0003-1262-6409}
\affiliation{%
  \institution{Dynatrace Research}
  \city{Linz}
  \country{Austria}
}

\author{Andreas Berger}
\email{andreas.berger@dynatrace.com}
\affiliation{%
  \institution{Dynatrace Research}
  \city{Linz}
  \country{Austria}
}

\begin{abstract}
  Rapidly-changing cloud environments that consist of heavily interconnected components are difficult to secure.
  Existing solutions often try to correlate many weak indicators to identify and reconstruct multi-step cyber attacks.
  The lack of a true, causal link between most of these indicators still leaves administrators with a lot of false-positives to browse through.
  We argue that cyber deception can improve the precision of attack detection systems, if used in a structured, and automatic way, i.e., in the form of so-called tripwires that ultimately span an attack graph, which assists attack reconstruction algorithms.
  This paper proposes an idea for a framework that combines cyber deception, automatic tripwire injection and attack graphs, which eventually enables us to reconstruct multi-step cyber attacks in modern cloud environments.
\end{abstract}

\ccsdesc[500]{Security and privacy~Intrusion/anomaly detection and malware mitigation}
\ccsdesc[300]{Security and privacy~Web application security}
\ccsdesc[100]{Security and privacy~Network security}

\keywords{multi-step cyber attacks,
    intrusion detection,
    cyber deception,
    cyber kill-chain,
    attack graphs,
    honeypots,
    honeytokens,
    tripwires
}

\maketitle


\section{Introduction}
Both academia and industry continue to work on systems that detect cyber attacks.
Promising results have been achieved for identifying vulnerabilities, malware, and malicious behavior, but, those systems still struggle to distinguish between mere anomalies and security-relevant incidents~\cite{kumarPracticalMachineLearning2017,sommerOutsideClosedWorld2010}.
Recent work focuses on correlating security alarms from various sources, to detect advanced threats that consist of multiple steps and evolve over long time periods~\cite{navarroSystematicSurveyMultistep2018}.
Especially in the domain of enterprise networks, such attacks are popularly known as Advanced Persistent Threats~(APTs)~\cite{chenStudyAdvancedPersistent2014} and are often modelled by a cyber kill-chain~\cite{hutchinsIntelligencedrivenComputerNetwork2011}. Since cloud threats cannot be represented well with this model we instead use the general term multi-step cyber attack.

Although these approaches yield more interpretable alarms, they still have troubles filtering out false-positives when applied in the real world~\cite{sommerOutsideClosedWorld2010}.
Furthermore, in the reconstruction of multi-step attacks, common approaches correlate IP addresses, alarm types, or time windows~\cite{navarroSystematicSurveyMultistep2018}, however, no approach fundamentally knows the true, causal link between alarms.

Securing cloud environments that consist of heavily interconnected components is particularly difficult.
Adversaries profit from the broadened attack surface and increased number of attack vectors.
Cyber deception, e.g., deploying honeypots, which are purposefully vulnerable entities that should attract adversaries, counteracts their unfair advantage by slowing them down~\cite{roweIntroductionCyberdeception2016}.
However, honeypots are often configured and deployed manually, making them scale poorly with rapidly-changing cloud environments.

We focus on three aspects to progress on these open issues:

\textbf{(1) Cyber Deception.}
To improve the precision of detection methods, we argue that it is important to have strong indicators of compromise~(IoCs), instead of correlating many weak indicators.
To achieve this, we introduce tripwires, which describe a deceptive scenario, and combine honeypots and honeytokens~\cite{spitznerHoneytokensOtherHoneypot2003}.

\textbf{(2) Automatic Tripwire Injection.}
To address the complex nature of cloud environments, tripwires are automatically, and strategically injected into existing applications.
Our framework manages their deployment and reacts to changes in the cloud environment.

\textbf{(3) Attack Graph Reconstruction.}
A tripwire consists of multiple, related deceptive components, e.g., a private SSH~key that acts as the lure, and the SSH~server that is the decoy.
Thereby, tripwires naturally form an attack graph~\cite{phillipsGraphbasedSystemNetworkvulnerability1998}, e.g., if the decoy is accessed, we know that the adversary must have gotten the keys first, which we deliberately placed on a different entity to trace the attack path.
Incoming alarms can then be projected onto the attack graph to reconstruct multi-step attacks.

To this end, we ask the following research question: \textit{``Are automatically injected tripwires suitable to reconstruct multi-step cyber attacks in modern cloud environments?''}.
This paper proposes an idea for a framework that can answer this question.


\section{Framework}
\begin{figure}[t]
  \centering
  \includegraphics[width=\columnwidth,page=1]{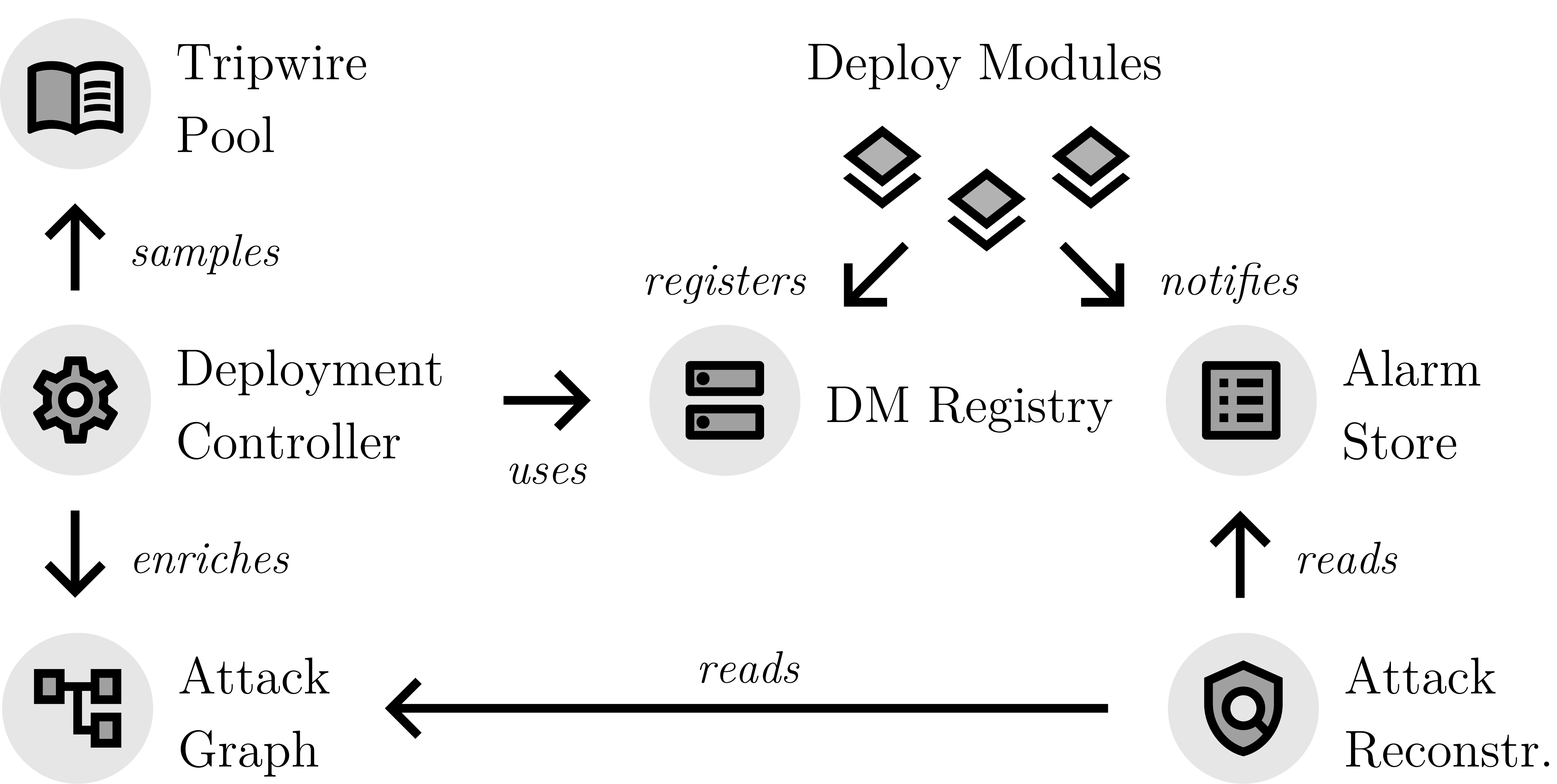}
  \vspace{-0.9em}
  \caption{The framework describes the life cycle of tripwires in cloud environments, from deployment, alarm and attack graph storage, to attack reconstruction.}
  \label{fig:controller}
\end{figure}

Our framework is designed for cloud environments, i.e., applications that are primarily exposed on the public internet.
Applications have in common that they often use libraries for common use cases such as web services or database communication.
We automatically inject tripwires into those libraries with the 7-component framework depicted in Figure~\ref{fig:controller}.
These components interact as follows:

\newpage

\textbf{Deploy Modules (DMs).}
Process hooks can identify used libraries and even inject code into applications at runtime.
Thereby, we can automatically inject deceptive components into applications, e.g., additional HTTP~endpoints on web applications.
Likewise, a process that runs on every host can deploy honeypots, or, create and modify files.
We call the component that injects lures or decoys, and optionally sets up an alarm system, a~DM.
The associated alarm system watches access attempts to the deployed lures or decoys.

\textbf{DM Registry (DMR).}
After scanning the cloud environment for injectable processes and hosts, many such DMs are instantiated.
Each of them registers itself in the DMR.

\begin{figure}[t]
  \centering
  \includegraphics[width=\columnwidth,page=2]{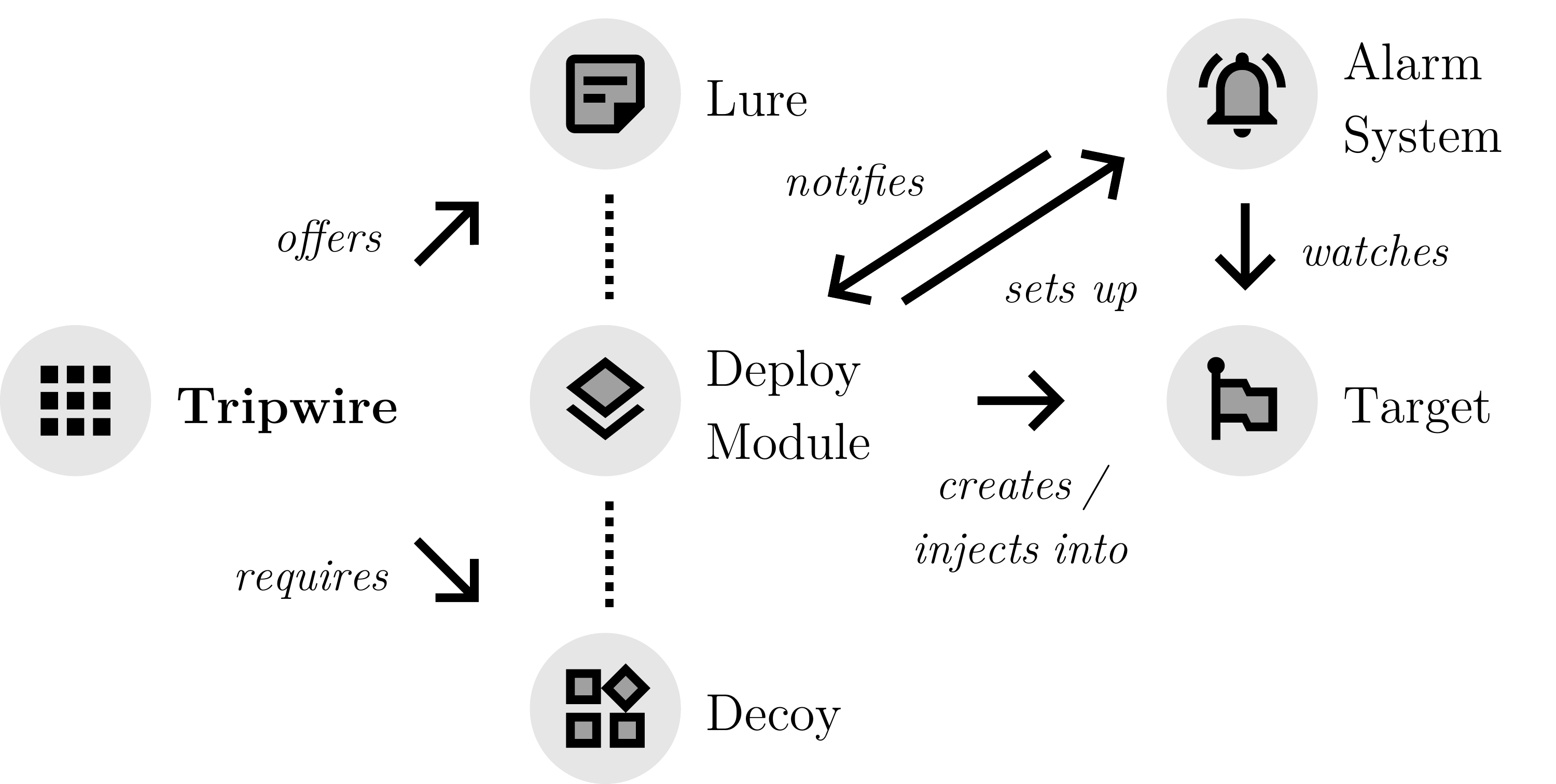}
  \caption{A tripwire describes the relation between lures, decoys, their deployment on some target via a deploy module, and its associated alarm system.}
  \vspace{-0.9em}
  \label{fig:tripwire}
\end{figure}

\textbf{Tripwire Pool (TP).}
A tripwire is a manually-created definition of a deceptive scenario.
Many instances of these are stored inside a~TP.
Take the AWS S3 bucket honeytoken\footnote{For an implementation, see \url{https://bitbucket.org/asecurityteam/spacecrab}} as a concrete example for the abstract model in Figure~\ref{fig:tripwire}.
There, we need at least two DMs: One creates an empty bucket (the target\footnote{The terms lure and decoy in Figure~\ref{fig:tripwire} solely mean an abstract concept of them. They do not represent a target. Eventually, a DM either creates a concrete instance, which then becomes a target, or, a DM injects something into an existing target (e.g., applications, files). The bucket tripwire specifies that it requires an empty bucket decoy to work. A~DM with access to the AWS~API will ultimately create that concrete decoy.}) and monitors accesses attempts to it, while one or more other DMs deploy the access token lure.
Various DMs allow to inject access token strings (e.g., into files, HTTP~headers, cookies, REST~endpoints).

\textbf{Deployment Controller (DC).}
The DC first queries all available DMs from the DMR and then strategically decides which DMs to use.
Its goal is to cover the entire cloud environment with deceptive components and react to re-deployments of applications.

\textbf{Attack Graph (AG).}
The DC also populates the AG to keep track of the deployment of deceptive components and their relationships, i.e., which lure is required to access which decoy.

\textbf{Alarm Store (AS).}
Every DM notifies the AS upon new alarms that were caught by an alarm system.
While the technical implementation might be different, we imagine that the AS stores alarms in a condensed form, which makes querying the AS more efficient.

\textbf{Attack Reconstruction (AR).}
The reconstruction algorithm takes the AS and the AG as its input, and is tasked to reconstruct attacks.
Given this data, we intend to use common backward and forward tracking algorithms~\cite{kingBacktrackingIntrusions2003,liuTimelyCausalityAnalysis2018} or variations of them to reconstruct multi-step cyber attacks.


\section{Conclusion}
We proposed an idea for a framework that strategically deploys deceptive components in modern cloud environments.
We described both the general framework, as well as tripwires, which are a definition of managed, adaptive, automatically injected, and strategically placed deceptive components, combining lures and decoys.

We envision that automatically injected deceptive components that form an attack graph offer a way to find true, causal links between alarms, thus improving the reconstruction of multi-step cyber attacks.
In future work, we plan to implement such a system and evaluate various attack reconstruction algorithms.

\begin{acks}
We thank all members of the Dynatrace research group and the anonymous reviewers for their valuable feedback on this work.
\end{acks}


\makeatletter\if@ACM@anonymous\newpage\fi\makeatother

\bibliographystyle{acmart}
\bibliography{eicc2020_kahlhofer_tripwires_poster}


\begin{thebibliography}{10}


\ifx \showCODEN    \undefined \def \showCODEN     #1{\unskip}     \fi
\ifx \showDOI      \undefined \def \showDOI       #1{#1}\fi
\ifx \showISBNx    \undefined \def \showISBNx     #1{\unskip}     \fi
\ifx \showISBNxiii \undefined \def \showISBNxiii  #1{\unskip}     \fi
\ifx \showISSN     \undefined \def \showISSN      #1{\unskip}     \fi
\ifx \showLCCN     \undefined \def \showLCCN      #1{\unskip}     \fi
\ifx \shownote     \undefined \def \shownote      #1{#1}          \fi
\ifx \showarticletitle \undefined \def \showarticletitle #1{#1}   \fi
\ifx \showURL      \undefined \def \showURL       {\relax}        \fi
\providecommand\bibfield[2]{#2}
\providecommand\bibinfo[2]{#2}
\providecommand\natexlab[1]{#1}
\providecommand\showeprint[2][]{arXiv:#2}

\bibitem[\protect\citeauthoryear{Chen, Desmet, and Huygens}{Chen
  et~al\mbox{.}}{2014}]%
        {chenStudyAdvancedPersistent2014}
\bibfield{author}{\bibinfo{person}{Ping Chen}, \bibinfo{person}{Lieven Desmet},
  {and} \bibinfo{person}{Christophe Huygens}.} \bibinfo{year}{2014}\natexlab{}.
\newblock \showarticletitle{A {{Study}} on {{Advanced Persistent Threats}}}. In
  \bibinfo{booktitle}{\emph{Communications and {{Multimedia Security}} ({{CMS}}
  '14)}} \emph{(\bibinfo{series}{LNCS}, Vol.~\bibinfo{volume}{8735})}.
  \bibinfo{publisher}{{Springer}}, \bibinfo{address}{{Berlin, Heidelberg}},
  \bibinfo{pages}{63--72}.
\newblock
\showISBNx{978-3-662-44885-4}
\urldef\tempurl%
\url{https://doi.org/10/gg33gx}
\showDOI{\tempurl}


\bibitem[\protect\citeauthoryear{Hutchins, Cloppert, and Amin}{Hutchins
  et~al\mbox{.}}{2011}]%
        {hutchinsIntelligencedrivenComputerNetwork2011}
\bibfield{author}{\bibinfo{person}{Eric Hutchins}, \bibinfo{person}{Michael
  Cloppert}, {and} \bibinfo{person}{Rohan Amin}.}
  \bibinfo{year}{2011}\natexlab{}.
\newblock \showarticletitle{Intelligence-Driven Computer Network Defense
  Informed by Analysis of Adversary Campaigns and Intrusion Kill Chains}. In
  \bibinfo{booktitle}{\emph{6th {{International Conference}} on {{Information
  Warfare}} and {{Security}}}} \emph{(\bibinfo{series}{{{ICIW}} '11})}.
  \bibinfo{publisher}{ACPI}, \bibinfo{address}{{Washington, DC}},
  \bibinfo{pages}{113--125}.
\newblock


\bibitem[\protect\citeauthoryear{King and Chen}{King and Chen}{2003}]%
        {kingBacktrackingIntrusions2003}
\bibfield{author}{\bibinfo{person}{Samuel~T. King} {and}
  \bibinfo{person}{Peter~M. Chen}.} \bibinfo{year}{2003}\natexlab{}.
\newblock \showarticletitle{Backtracking Intrusions}. In
  \bibinfo{booktitle}{\emph{Proceedings of the Nineteenth {{ACM}} Symposium on
  {{Operating}} Systems Principles}} \emph{(\bibinfo{series}{{{SOSP}} '03})}.
  \bibinfo{publisher}{ACM}, \bibinfo{address}{{Bolton Landing, NY}},
  \bibinfo{pages}{223--236}.
\newblock
\showISBNx{978-1-58113-757-6}
\urldef\tempurl%
\url{https://doi.org/10/b47kcm}
\showDOI{\tempurl}


\bibitem[\protect\citeauthoryear{Kumar, Wicker, and Swann}{Kumar
  et~al\mbox{.}}{2017}]%
        {kumarPracticalMachineLearning2017}
\bibfield{author}{\bibinfo{person}{Ram Shankar~Siva Kumar},
  \bibinfo{person}{Andrew Wicker}, {and} \bibinfo{person}{Matt Swann}.}
  \bibinfo{year}{2017}\natexlab{}.
\newblock \showarticletitle{Practical {{Machine Learning}} for {{Cloud
  Intrusion Detection}}: {{Challenges}} and the {{Way Forward}}}. In
  \bibinfo{booktitle}{\emph{Proceedings of the 10th {{ACM Workshop}} on
  {{Artificial Intelligence}} and {{Security}}}}
  \emph{(\bibinfo{series}{{{AISec}} '17})}. \bibinfo{publisher}{ACM},
  \bibinfo{address}{{Dallas, Texas}}, \bibinfo{pages}{81--90}.
\newblock
\showISBNx{978-1-4503-5202-4}
\urldef\tempurl%
\url{https://doi.org/10/ggkqcp}
\showDOI{\tempurl}


\bibitem[\protect\citeauthoryear{Liu, Zhang, Li, Jee, Li, Wu, Rhee, and
  Mittal}{Liu et~al\mbox{.}}{2018}]%
        {liuTimelyCausalityAnalysis2018}
\bibfield{author}{\bibinfo{person}{Yushan Liu}, \bibinfo{person}{Mu Zhang},
  \bibinfo{person}{Ding Li}, \bibinfo{person}{Kangkook Jee},
  \bibinfo{person}{Zhichun Li}, \bibinfo{person}{Zhenyu Wu},
  \bibinfo{person}{Junghwan Rhee}, {and} \bibinfo{person}{Prateek Mittal}.}
  \bibinfo{year}{2018}\natexlab{}.
\newblock \showarticletitle{Towards a {{Timely Causality Analysis}} for
  {{Enterprise Security}}}. In \bibinfo{booktitle}{\emph{Proceedings 2018
  {{Network}} and {{Distributed System Security Symposium}}}}
  \emph{(\bibinfo{series}{{{NDSS}} '18})}. \bibinfo{publisher}{{Internet
  Society}}, \bibinfo{address}{{San Diego, CA}}.
\newblock
\showISBNx{978-1-891562-49-5}
\urldef\tempurl%
\url{https://doi.org/10/ggk6gj}
\showDOI{\tempurl}


\bibitem[\protect\citeauthoryear{Navarro, Deruyver, and Parrend}{Navarro
  et~al\mbox{.}}{2018}]%
        {navarroSystematicSurveyMultistep2018}
\bibfield{author}{\bibinfo{person}{Julio Navarro}, \bibinfo{person}{Aline
  Deruyver}, {and} \bibinfo{person}{Pierre Parrend}.}
  \bibinfo{year}{2018}\natexlab{}.
\newblock \showarticletitle{A Systematic Survey on Multi-Step Attack
  Detection}.
\newblock \bibinfo{journal}{\emph{Computers \& Security}}  \bibinfo{volume}{76}
  (\bibinfo{date}{July} \bibinfo{year}{2018}), \bibinfo{pages}{214--249}.
\newblock
\showISSN{0167-4048}
\urldef\tempurl%
\url{https://doi.org/10/gdv95j}
\showDOI{\tempurl}


\bibitem[\protect\citeauthoryear{Phillips and Swiler}{Phillips and
  Swiler}{1998}]%
        {phillipsGraphbasedSystemNetworkvulnerability1998}
\bibfield{author}{\bibinfo{person}{Cynthia Phillips} {and}
  \bibinfo{person}{Laura~Painton Swiler}.} \bibinfo{year}{1998}\natexlab{}.
\newblock \showarticletitle{A Graph-Based System for Network-Vulnerability
  Analysis}. In \bibinfo{booktitle}{\emph{Proc. of the 1998 Workshop on {{New}}
  Security Paradigms}} \emph{(\bibinfo{series}{{{NSPW}} '98})}.
  \bibinfo{publisher}{ACM}, \bibinfo{address}{{New York, NY}},
  \bibinfo{pages}{71--79}.
\newblock
\showISBNx{978-1-58113-168-0}
\urldef\tempurl%
\url{https://doi.org/10/cxxqrd}
\showDOI{\tempurl}


\bibitem[\protect\citeauthoryear{Rowe and Rrushi}{Rowe and Rrushi}{2016}]%
        {roweIntroductionCyberdeception2016}
\bibfield{author}{\bibinfo{person}{Neil~C. Rowe} {and} \bibinfo{person}{Julian
  Rrushi}.} \bibinfo{year}{2016}\natexlab{}.
\newblock \bibinfo{booktitle}{\emph{Introduction to {{Cyberdeception}}}}.
\newblock \bibinfo{publisher}{Springer International Publishing},
  \bibinfo{address}{{Cham}}.
\newblock
\showISBNx{978-3-319-41185-9 978-3-319-41187-3}
\urldef\tempurl%
\url{https://doi.org/10/d65q}
\showDOI{\tempurl}


\bibitem[\protect\citeauthoryear{Sommer and Paxson}{Sommer and Paxson}{2010}]%
        {sommerOutsideClosedWorld2010}
\bibfield{author}{\bibinfo{person}{Robin Sommer} {and} \bibinfo{person}{Vern
  Paxson}.} \bibinfo{year}{2010}\natexlab{}.
\newblock \showarticletitle{Outside the {{Closed World}}: {{On Using Machine
  Learning}} for {{Network Intrusion Detection}}}. In
  \bibinfo{booktitle}{\emph{2010 {{IEEE Symposium}} on {{Security}} and
  {{Privacy}}}} \emph{(\bibinfo{series}{S\&{{P}} '10})}.
  \bibinfo{publisher}{IEEE Computer Society}, \bibinfo{address}{{Oakland, CA}},
  \bibinfo{pages}{305--316}.
\newblock
\showISSN{1081-6011}
\urldef\tempurl%
\url{https://doi.org/10/cgp43q}
\showDOI{\tempurl}


\bibitem[\protect\citeauthoryear{Spitzner}{Spitzner}{2003}]%
        {spitznerHoneytokensOtherHoneypot2003}
\bibfield{author}{\bibinfo{person}{Lance Spitzner}.}
  \bibinfo{year}{2003}\natexlab{}.
\newblock \bibinfo{title}{Honeytokens: {{The Other Honeypot}}}.
\newblock
\newblock
\urldef\tempurl%
\url{https://www.symantec.com/connect/articles/honeytokens-other-honeypot}
\showURL{%
\tempurl}


\end{thebibliography}

\end{document}